\documentclass[12pt]{article}

\usepackage{amssymb}
\usepackage{amsfonts}
\usepackage{amsmath}
\usepackage[mathscr]{eucal}
\usepackage{amssymb}
\usepackage{amsthm}
\usepackage{mathrsfs}
\usepackage{bbold}
\usepackage{bm}
\usepackage{graphicx}
\usepackage{caption}
\usepackage[english]{babel}
\usepackage[T2A]{fontenc}
\usepackage[utf8]{inputenc}
\usepackage{cite}

\def\theequation{\arabic{section}.\arabic{equation}}
\numberwithin{equation}{section}

\newcommand{\be}{\begin{equation}}
\newcommand{\ee}{\end{equation}}
\newcommand{\bea}{\begin{eqnarray}}
\newcommand{\eea}{\end{eqnarray}}

\newcommand{\p}[1]{(\ref{#1})}

\begin{document}

\begin{titlepage}

\vspace*{0.7cm}

\begin{center}

{\LARGE\bf Light-front description of infinite spin fields

\vspace{0.2cm}

in six-dimensional Minkowski space}

\vspace{1.5cm}

{\large\bf I.L.\,Buchbinder$^{1,2,3,4}$\!\!,\ \ \
S.A.\,Fedoruk$^4$\!\!,\ \ \ A.P.\,Isaev$^{4,5}$}

\vspace{1.0cm}

\ $^1${\it Center of Theoretical Physics,
Tomsk State Pedagogical University, \\
634041 Tomsk, Russia}, \\
{\tt joseph@tspu.edu.ru}

\vskip 0.3cm

\ $^2${\it National Research Tomsk State  University,\\634050 Tomsk,
Russia}

\vskip 0.3cm

\ $^3${\it Lab of Theor. Cosmology, International Center of Gravity and Cosmos, \\
Tomsk State University of Control Systems and
Radioelectronics (TUSUR), \\ 634050, Tomsk, Russia}

\vskip 0.3cm

\ $^4${\it Bogoliubov Laboratory of Theoretical Physics,
Joint Institute for Nuclear Research, \\
141980 Dubna, Moscow Region, Russia}, \\
{\tt fedoruk@theor.jinr.ru, isaevap@theor.jinr.ru}

\vskip 0.3cm

\ $^5${\it Faculty of Physics, Lomonosov Moscow State University,
\\ 119991 Moscow, Russia}

\end{center}

\vspace{1.5cm}

\begin{abstract}
We present a new $6D$ infinite spin field theory in the light-front formulation.
The Lorentz-covariant counterparts of these fields depend on 6-vector coordinates and additional spinor variables.
Casimir operators in this realization are found.
We obtain infinite-spin fields in the light-cone frame which depend on two sets of the $\mathrm{SU}(2)$-harmonic variables.
The generators of the $6D$ Poincar\'e group and the infinite spin field action in the light-front formulation are presented.
\end{abstract}

\vspace{1.5cm}


\phantom{Keywords: }

\vspace{1cm}

\end{titlepage}

\setcounter{footnote}{0}
\setcounter{equation}{0}

\newpage

\setcounter{equation}0
\section{Introduction}

The study of various aspects of classical and quantum field theory
in higher dimensions attracts attention basically due to connections
with the low-energy limit of superstring theory and miraculous
cancelations of some divergences in supersymmetric field models. One
of such aspects is a description of the massless representations of
the Poincar\'e group in multi-dimensional spaces (see e.g. the
recent works \cite{W,KUZ,BFIP,BFI-21})\footnote{For earlier activity in this direction see
e.g. \cite{BB1,BB2} and the references therein.}. In this
paper, we continue our study of field irreducible massless
representations of the six-dimensional Poincar\'e group \cite{BFIP,BFI-21} focusing on the infinite spin representations and
their Lagrangian formulation.

The study of the infinite spin representations of the Poincar\'e
group \cite{Wigner39,Wigner47,BargWigner}, their field realizations and dynamical description aroused
considerable interest, which led to the formation of a certain
research branch mainly in the context of the theory of higher spin
fields (see e.g. the review \cite{BekSk} and earlier references
therein, and recent papers \cite{BekMou,Bekaert:2017xin,Najafizadeh:2017tin,HabZin,AlkGr,Metsaev18,BFIR,
BuchKrTak,BuchIFKr,ACG18,Metsaev18a,BFI,Metsaev19,BKSZ,MN20,MN22,BuchIFKr22}) where
aspects of interactions and supersymmetry of infinite spin fields
have been examined). Since field realizations of the Poincar\'e
group representations in each concrete dimension have specific
features, infinite spin fields in higher dimensions deserve a separate
study.

To construct infinite spin fields in six-dimensional Minkowski
space, we should describe a possible spectrum of states corresponding
to these fields and, first of all, to clarify the spin structure. It can be
achieved by considering massless representations of the $6D$
Poincar\'e group in state space in terms of the canonically
conjugate position and momentum operators, as well as the canonically
conjugate pair of spinor operators. Irreducible representations are
formulated in terms of second-, fourth- and sixth-order Casimir
operators, respectively. The corresponding eigenvalues for these
Casimir operators in the irreducible infinite spin representation are
$0\,,-\mu^2\,,-\mu^{2}s(s+1)$ respectively, where $\mu$ is a
nonzero real parameter and $s$ is a non-negative integer or
half-integer number (see the details in \cite{BFIP}).

To describe the spin structure of the $6D$ infinite spin fields, it is
natural to refer to the light-cone frame for massless fields,
where the eigenvalues of the energy-momentum operator are
$p^0=p^5=k$, $p^{\hat a}=0$, ${\hat a}=1,2,3,4\,$ with some nonzero
real parameter $k$. Here a remarkable result was unexpectedly discovered that any infinite spin field in this frame is
necessarily a function on bi-harmonic space with the harmonics
$u^{\pm}\,,v^{\pm}$ which were earlier essentially used to construct
the unconstrained superfield formulation of $4D\,, {\cal N}=2$
supersymmetric field theories \cite{GIKOS,GIOS}.  Taking into
account this result, it is natural then to go to the light-front
coordinate system $x^{\pm},x^{\hat a}\,, a=1,2,3,4\,$, which
inherits the properties of the light-cone frame \cite{Dirac49}. Thus,
we arrive at the function of both $x^{\pm},x^{\hat a}$ and harmonics
$u^{\pm}\,,v^{\pm}$ which is considered as the infinite spin field
in the light-front coordinate system. The field dynamics
in the light-front coordinates can be constructed following the generic
scheme \cite{Dirac49} (see also
\cite{Bengtsson2Brink,Bengtsson2Linden,Sieg88,Sieg99,Metsaev05,PoSk}).

The paper is organized as follows. In Sect.\,2, we discuss the
description of irreducible infinite spin representations of the $6D$
Poincar\'e group in state space formulated in terms of the
position and momentum operators and spin operators. These operators
are $6D$ vectors and a pair of $\mathrm{SU}(2)$ Majorana-Weyl
spinors. We find expressions of the fourth- and sixth-orders Casimir
operators for the system under consideration and discuss the
conditions leading to fixing the eigenvalues of these operators on
physical states. In Sect.\,3, we derive infinite spin fields in
the light-cone frame. Here we show that these fields are the function on
bi-harmonic space with two sets of $\mathrm{SU}(2)$
harmonics $v_{i}^\pm$ and $v_{\underline{i}}^\pm$. The harmonics
obtained here describe the coset space
$[\mathrm{SU}(2)\,{\otimes}\,\mathrm{SU}(2)]/\mathrm{U}(1)$. Such a
harmonic field possesses a harmonic charge which is determined by
the eigenvalue of the sixth-order Casimir operator. We describe the
general structure of the harmonic field in the light-cone frame and show
that it is given by an infinite expansion in the harmonics. Using these
results, in Sect.\,4, we develop the light-front dynamical formulation
of an infinite spin field.  We find the generators of the $6D$
Poincar\'e group for the fields under consideration and propose the
corresponding action. An important point in this approach is the use
of harmonics as additional coordinates, which greatly simplifies the
field analysis. In Sect.\,5, we summarize the results obtained.
Appendix\,A  is devoted to the calculation of the sixth-order Casimir
operator for the system considered. In Appendix\,B, we find the spinor
part of the $6D$ Lorentz algebra generators.

\setcounter{equation}0
\section{Irreducible massless representation of the $D6$ Poincar\'e group}

In this section, we discuss the construction of a massless
irreducible representation of the six-dimensional Poincar\'e group
emphasizing the specific use of spinor operators.

We consider the representations in the space of states described by vectors
$|\Psi\rangle \,$. The basic operators acting in this space are
\begin{equation}
\label{op}
x^{a}\,,\quad p_{a}\,;\qquad \xi_\alpha^I\,,\quad \rho^{\alpha I}\,.
\end{equation}
Here the Hermitian coordinate $x^{a}=(x^{a})^\dagger$ and momentum $p_{a}=(p_{a})^\dagger$ operators are
components of the six-vectors, $a=0,1,\ldots,5$ and they obey the standard commutation relations
\begin{equation}
\label{x-p}
[x^{a},p_{b}]= i\delta^{a}_{b}\,.
\end{equation}
The operators $\xi_\alpha^I$, $\rho^{\alpha I}$ are the
$\mathrm{SU}(2)$ Majorana-Weyl spinors\footnote {We use the spinor
conventions of the works \cite{BFIP,BFI-21}.}, where
$\alpha=1,2,3,4$ and $I=1,2$ are, respectively, the spinorial
$\mathrm{SU}^*(4)$ and internal $\mathrm{SU}(2)$ indices. The
Hermitian conjugation for these operators is defined as follows:
\begin{equation}
\label{xi}
(\xi_\alpha^I)^\dagger =\epsilon_{IJ}B_{\dot\alpha}{}^\beta \xi_\beta^J\,,\qquad
(\rho^{\alpha I})^\dagger = \epsilon_{IJ}\rho^{\beta J}(B^{-1})_\beta{}^{\dot\alpha}\,,
\end{equation}
where $B_{\dot\alpha}{}^\beta$ is the matrix related to complex
conjugation, and the antisymmetric tensors $\epsilon_{IJ}$,
$\epsilon^{IJ}$ have the components $\epsilon_{12}=\epsilon^{21}=1$ (see
\cite{BFI-21} for details). Nonzero commutation relations for the
operators $\xi_\alpha^I$, $\rho^{\alpha I}$ have the form
\begin{equation}
\label{com-sp}
\left[ \xi_\alpha^I\,, \rho^\beta_J\right] = i\delta_\alpha^\beta \delta^I{}_J \,.
\end{equation}
The operators $\xi_\alpha^I$, $\rho^{\alpha I}$ are to describe the
spin degrees of freedom. The state space will be specified in the
next section.

We assume that the operators $p_a$ generate space-time translations.
In this case, the generators $\{P_a, M_{ab} \}$ of the algebra $\mathfrak{iso}(1,5)$ of the Poincar\'e group are realized as
\begin{eqnarray}
\label{P-gen}
P_a&=&  p_a \,, \\ [7pt]
\label{M-gen}
M_{ab}&=&
p_{a}x_{b}-p_{b}x_{a}+S_{ab} \,,
\end{eqnarray}
where the spin part of the Lorentz group generators looks like
\begin{equation}
\label{M-1}
S_{ab} = \xi_\alpha^I (\tilde\sigma_{ab})^{\alpha}{}_{\beta}\rho^\beta_I
= -\rho^\alpha_I (\sigma_{ab})_{\alpha}{}^{\beta}  \xi_\beta^I\,.
\end{equation}

We consider the massless representations where the quadratic Casimir operator $C_2 = P^2 = P^{a}P_{a}$
of the algebra $\mathfrak{iso}(1,5)$ has zero eigenvalues
\begin{equation}
\label{P2-0}
p^{a}p_{a}|\Psi\rangle =  0\,.
\end{equation}
In this case, the projection of the fourth-order Casimir operator in subspace \p{P2-0} has the form \cite{BFIP}
\begin{equation}
\label{C4-def}
C_4= \Pi^{a} \Pi_{a}  \, ,
\end{equation}
where
\begin{equation}
\label{Pi-def}
\Pi_a = P^{b}  M_{ba}  \, .
\end{equation}
As a result, we can see that in the representation \p{P-gen}, \p{M-gen}
and under the condition \p{P2-0} the operator \p{C4-def} takes the
following form:
\begin{equation}
\label{C4-expr}
C_4= - \,\tilde\ell\,\ell  \,,
\end{equation}
where the scalar operators $\ell$, $\tilde\ell$ are defined by the relations
\begin{equation}
\label{l-expr}
\ell:=\frac12\, \rho^\alpha_I(p_a\sigma^{a})_{\alpha\beta}\rho^\beta{}^I\,,\qquad
\tilde\ell:=\frac12\,  \xi_\alpha^I(p_a\tilde\sigma^{a})^{\alpha\beta}\xi_\beta{}_I\,.
\end{equation}
When deriving expression \p{C4-expr}, we used the relation
\p{sigma-id} for the $6D$ $\sigma$-matrices. The algebra of
operators \p{l-expr} is written in the form
\begin{equation}
\label{l-com}
[\tilde\ell,\ell]=N\, p^a p_a\,,
\end{equation}
where the operator $N$ is defined by the anticommutator
\begin{equation}
\label{N-def}
N:=\frac{i}{2}\, \{\xi_\alpha^I,\rho^\alpha_I\} \,.
\end{equation}
Besides, the operators \p{l-expr} are the Poincar\'e group
invariants and hence they commute with the generators \p{P-gen},
\p{M-gen}
\begin{equation}
\label{l-PM}
[P_a,\ell]=[P_a,\tilde\ell]=0\,\qquad [M_{ab},\ell]=[M_{ab},\tilde\ell]=0\,.
\end{equation}

The infinite spin representation is characterized by the condition that the fourth-order Casimir
operator has nonzero negative eigenvalue
\begin{equation}
\label{C4-const}
C_4\,|\Psi\rangle \ = \ -\,\mu^2\,|\Psi\rangle \,,
\end{equation}
where $\mu \neq 0$ is the dimensional real parameter which can be taken
positive $\mu \in \mathbb{R}_{> 0}$ without loss of generality. Using relations \p{C4-const} and \p{C4-expr}, we can see that
it is sufficient to define infinite spin states by the constraints
\begin{equation}
\label{C4-constr}
\ell\,|\Psi\rangle \ = \ \mu\,|\Psi\rangle \,,\qquad \tilde\ell\,|\Psi\rangle \ = \ \mu\,|\Psi\rangle \, .
\end{equation}

For massless representations \p{P2-0}, the sixth-order Casimir operator has the form \cite{BFIP}
\begin{equation}
\label{C6-Casimir}
C_6=
-\, \Pi^b M_{ba}\, \Pi_c M^{ca}
\ + \ \frac12\,\Big(M^{ab}M_{ab}-8\Big)\, \Pi^{a} \Pi_{a} \,,
\end{equation}
where the operator $\Pi_a$ is defined in \p{Pi-def}.
In the representation \p{P-gen}, \p{M-gen} and under the conditions \p{P2-0}, \p{C4-const}, we obtain\footnote{See the details in Appendix A.}
\begin{equation}
\label{C6-Cas}
C_6\,|\Psi\rangle \ = \ -\, \mu^2\, J_{\mathrm{i}}J_{\mathrm{i}}\,|\Psi\rangle\,,
\end{equation}
where the operators $J_{\mathrm{i}}$ $(\mathrm{i}=1,2,3)$ are defined as follows:
\begin{equation}\label{Ta-def}
J_{\mathrm{i}} \ := \ \frac{i}{2}\,\xi_\alpha^I(\sigma_{\mathrm{i}})_I{}^J \rho^\alpha_J\,.
\end{equation}
Here $\sigma_{\mathrm{i}}$ are the Pauli matrices. The operators $J_{\mathrm{i}}$ form the $\mathfrak{su}(2)$ algebra
\begin{equation}\label{Ta-alg}
[J_{\mathrm{i}},J_{\mathrm{j}}]= i\epsilon_{{\mathrm{i}}{\mathrm{j}}{\mathrm{k}}}J_{\mathrm{k}}\,.
\end{equation}
Expression \p{C6-Cas} for the operator $C_6$ is the same as in
\cite{BFIP} but the realization of the generators $J_{\mathrm{i}}\in
\mathfrak{su}(2)$ in \cite{BFIP} is different.

As it was shown in \cite{BFIP}, the space $V$ of irreducible infinite
spin representation is induced from the space of finite dimensional representation of \p{Ta-def} and the operator $C_6$ acts as follows:
\begin{equation}
\label{C6-ir}
C_6\,|\Psi\rangle \ = \ -\,\mu^2\, s(s+1)\,|\Psi\rangle\,,
\end{equation}
where $s$ is a nonzero integer or half-integer number, $s \in \mathbb{Z}_{\geq 0}/2$. Therefore,
the states corresponding to the infinite spin irreducible representation obey the constraints
\begin{equation}\label{eq-3}
J_{\mathrm{i}}J_{\mathrm{i}}\,|\Psi\rangle \ = \ s(s+1)\,|\Psi\rangle \,,
\end{equation}
where the operators  $J_{\mathrm{i}}$ are defined in \p{Ta-def}.

Note that the $\mathfrak{su}(2)$ algebra generators \p{Ta-def} commute with the $\mathrm{SU}(2)$ scalar
operators \p{l-expr}:
\begin{equation}
\label{su2-l} [J_{\mathrm{i}},\ell]=[J_{\mathrm{i}},\tilde\ell]=0
\,.
\end{equation}
Besides, the operators \p{Ta-def} commute with generators of six-dimensional translations \p{P-gen} and
with the Lorentz algebra $\mathfrak{so}(1,5)$ generators \p{M-gen}, \p{M-1}:
\begin{equation}
\label{M-J}
[P_{a},J_{\mathrm{i}}]=0\,,\qquad [M_{ab},J_{\mathrm{i}}]= [S_{ab},J_{\mathrm{i}}]=0\,.
\end{equation}

It is worth noting that the algebra
$\mathfrak{so}(1,5)=\mathfrak{su}^*(4)$ generated by \p{M-1} is dual
to the algebra $\mathfrak{su}(2)$ with the generators \p{Ta-def} in
the sense of Howe duality \cite{Howe}.

It was shown earlier \cite{BekMou,BB1,BB2} that in the vector
approach an infinite spin representation of $\mathfrak{so}(1,5)$
requires the use of the 6-dimensional Heisenberg algebra \p{x-p}
generated by the operators of position $x^a$ and momentum $p_a$ and two
additional 6-dimensional Heisenberg algebras with the coordinate
operators $y_1^a$, $y_2^a$ and their momentum operators
$p^{(y)}_{1a}$, $p^{(y)}_{2a}$. On the other hand, we proved in
\cite{BFI-21} that, in the twistor formulation, the
$\mathfrak{iso}(1,5)$ representations of infinite spin are
necessarily described in the bi-twistor space, which is defined by
two pairs of canonically-conjugated $\mathrm{SU}(2)$ Majorana–Weyl
spinors of type \p{com-sp} (but having nonzero mass dimensions).
Here we have shown that $6D$ infinite spin representations can be
described in the space defined by only one 6-dimensional Heisenberg
algebra \p{x-p} and one pair of canonically conjugated
$\mathrm{SU}(2)$ Majorana–Weyl spinors \p{com-sp}, as it was
indicated in  \p{op}.

In the next section, we will describe the infinite spin vectors $|\Psi\rangle$ in terms of appropriate fields.

\setcounter{equation}0
\section{Infinite spin fields in the light-cone frame}

We consider a structure of $D6$ infinite spin fields in the light-cone frame. This frame is defined
by the following conditions on eigenvalues of the energy-momentum operator:
\begin{equation}
\label{P-st}
p^0=p^5=k\,,\qquad p^{\hat a}=0\,,\qquad {\hat a}=1,2,3,4\,,
\end{equation}
where $k$ is a nonzero real parameter of mass dimension.
Using the light-cone coordinates
$p^\pm=\left(p^0\pm p^5\right)/{\sqrt{2}}$, one gets for the same frame
\begin{equation}
\label{P-st-lc}
p^+=\sqrt{2}k\,,\qquad p^-=p^{\hat a}=0\,.
\end{equation}

Consider the $\mathrm{SU}^*(4)$ spinors $\xi_\alpha^I$, $\rho^{\alpha I}$ with a four-component spinor index
$\alpha$ and present them as objects with two-component indices as follows:
\begin{equation}
\label{pi-2-2}
\xi_\alpha^I=(\xi_i^I,\xi_{\underline{i}}^I)\,,\qquad\quad \rho^\alpha_I=(\rho^i_I,\rho^{\underline{i}}_I)\,,
\end{equation}
where the two-component indices take the values
$i=1,2$ and $\underline{i}=1,2$, i.e. $i=\alpha$ for $\alpha=1,2$, and
 $\underline{i}=\alpha-2$ for $\alpha=3,4$.

In the light-cone frame \p{P-st} the operators  \p{l-expr} take the form
\begin{equation}
\label{l-tl}
\tilde\ell \ = \  k\,\epsilon^{ij}\epsilon_{IJ}\,\xi_i^I \xi_j^J\,,\qquad\quad
\ell \ = \ k\,\epsilon_{\underline{i}\underline{j}}\epsilon^{IJ}\,\rho^{\underline{i}}_I \rho^{\underline{j}}_J\,,
\end{equation}
where the matrices $\tilde\sigma^-$ \p{t-sigma+-} and $\sigma^-$ \p{sigma+-} were used.
Then in this frame the constraints
\begin{equation}
\label{C4-constr-1}
\tilde\ell= \mu \, , \qquad \ell= \mu
\end{equation}
from \p{C4-constr} are written in the form
\begin{equation}
\label{harm-cond-1}
\epsilon_{IJ}u_i^I u_j^J=\epsilon_{ij}\,,\qquad\quad \epsilon_{IJ}v_{\underline{i}}^I v_{\underline{j}}^J=\epsilon_{\underline{i}\underline{j}}\,,
\end{equation}
where we
have used the spinor variables
\begin{equation}
\label{harm-1}
u_i^I:= \sqrt{2k/\mu}\; \xi_i^I\,;\qquad\quad v_{\underline{i}}^I:= \sqrt{2k/\mu}\; \rho_{\underline{i}}^I\,,\qquad \rho_{\underline{i}}^I
=\epsilon_{\underline{i}\underline{j}}\epsilon^{IJ}\rho^{\underline{j}}_J\,.
\end{equation}
The conditions \p{xi} in terms of the $\mathrm{SU}(2)$ spinors
\p{harm-1} look like
\begin{equation}
\label{u-real}
(u_i^I)^* =-\epsilon_{IJ}\epsilon^{ij}u_j^J\,, \qquad\quad (v_{\underline{i}}^I)^* =-\epsilon_{IJ}\epsilon^{\underline{i}\underline{j}}v_{\underline{j}}^J\,.
\end{equation}
Conditions \p{harm-cond-1}, \p{u-real} are nothing but the ones of unimodularity, $\det u=1$, $\det v=1$, and
unitarity, $u^\dagger u=1$, $v^\dagger v=1$, of the $2{\times}2$ matrices
\begin{equation}
\label{u-matrix}
u \ := \  \parallel\! u_i{}^I \!\parallel\,, \qquad v \ := \  \parallel\! v_{\underline{i}}{}^I \!\parallel\,.
\end{equation}
As a result, in the light-cone frame the variables $u_i^I$ and $v_{\underline{i}}^I$ \p{harm-1} are the elements of the $\mathrm{SU}(2)$ groups
and parameterize the compact space. Further, analogously to \cite{GIKOS,GIOS}, we will use the following notation:
\begin{equation}
\label{harm}
u_i^1=u_i^+\,,\quad  u_i^2=u_i^-\,,\qquad\quad v_{\underline{i}}^1=v_{\underline{i}}^+\,,
\quad v_{\underline{i}}^2=v_{\underline{i}}^-\,.
\end{equation}
In this notation, relations \p{harm-cond-1} are rewritten in the form \footnote{Note that the
$\mathrm{U}(1)$ charges $\pm$ in the variables $u^{\pm}$ and
$v^{\pm}$ have a different meaning than the light-cone indices
$\pm$ in quantities $p^{\pm}$. The later are the $SO(1,1)$ vector
indices.}
\begin{equation}
\label{harm-cond-1a}
u_i^+ u_j^- - u_j^+ u_i^-=\epsilon_{ij}\,,\qquad\quad
v_{\underline{i}}^+ v_{\underline{j}}^- - v_{\underline{j}}^+ v_{\underline{i}}^-=\epsilon_{\underline{i}\underline{j}}
\end{equation}
or, in the equivalent form
\begin{equation}
\label{harm-cond-2b}
u^i{}^+ u_i^- =1\,,\qquad\quad
v^{\underline{i}}{}^+ v_{\underline{i}}^- =1\,,
\end{equation}
where $u^i{}^\pm=\epsilon^{ij}u_j^\pm$, $u^{\underline{i}}{}^\pm=\epsilon^{\underline{i}\underline{j}}u_{\underline{j}}^\pm$.
Note that both $u^{\pm}$ and $v^{\pm}$ have the same indices $\pm$,
since they are obtained from the common $\mathrm{SU}(2)$-index $I$
for the $\mathrm{SU}(2)$ Majorana-Weyl spinors \p{pi-2-2}.

We emphasize that the variables $u^{\pm}$ and $v^{\pm}$ introduced in
\p{harm-1} completely determine the operators $\ell$, $\tilde\ell$ in \p{l-tl} and represent half of the different
canonical pairs in the algebra \p{com-sp}. We treat the second half
of the operators in \p{com-sp} as differential operators. Thus, one
considers a representation
where the operators
$\rho^i_I$ and $\xi_{\underline{i}}^I$ in the algebra \p{com-sp} are
realized as differential operators
\begin{equation}
\label{vw-dif}
\rho^i_I=-i\frac{\partial}{\partial \xi_i^I}=-i\sqrt{2k/\mu}\,\frac{\partial}{\partial u_i^I} \,,\qquad
\xi_{\underline{i}}^I=i\frac{\partial}{\partial \rho^{\underline{i}}_I}=
i\sqrt{2k/\mu}\,\epsilon_{ij}\epsilon^{IJ}\frac{\partial}{\partial v_j^J} \,.
\end{equation}
In the representation chosen, the operators $\ell$ and $\tilde\ell$
are realized by operators of multiplication by the functions of
$u^\pm$, $v^\pm$.

In such a representation, the $\mathfrak{su}(2)$-generators
$J_\pm:=J_1\pm i J_2$ and $J_3$, given by \p{Ta-def}, are written as
follows
\begin{equation}
\label{Ja-ab3}
J_\pm = D^{\pm\pm}_u +D^{\pm\pm}_v\,, \qquad
J_3 =\frac{1}{2}\left(D^{0}_u + D^{0}_v\right)\,,
\end{equation}
where
\begin{eqnarray}
\label{D-u}
D^{\pm\pm}_u:= u_i^\pm \frac{\partial}{\partial u_i^\mp}\,,&&
D^{0}_u:=u_i^+ \frac{\partial}{\partial u_i^+} - u_i^- \frac{\partial}{\partial u_i^-}\,, \\ [6pt]
\label{D-w}
D^{\pm\pm}_v:= v_{\underline{i}}^\pm \frac{\partial}{\partial v_{\underline{i}}^\mp}\,,&&
D^{0}_v:=v_{\underline{i}}^+ \frac{\partial}{\partial v_{\underline{i}}^+} - v_{\underline{i}}^- \frac{\partial}{\partial v_{\underline{i}}^-}
\end{eqnarray}
coincide with the harmonic derivatives in the notation \cite{GIKOS,GIOS}.

Relations \p{Ta-alg} are written in the form $[J_+,J_-]=2J_3$,
$[J_3,J_\pm]=\pm J_\pm$, and the Casimir operator of the
$\mathfrak{su}(2)$ algebra has the standard expression
$$
J_{\mathrm{i}}J_{\mathrm{i}} \ = \ J_3(J_3+1)+J_-J_+  \ = \  -J_3(-J_3+1)+J_+J_-\,.
$$
As a solution to the irreducibility condition \p{C6-Cas}, \p{C6-ir} for $6D$ representations
we take the highest weight vector $|\Psi^{(2s)}\rangle$
which is defined by the equations
\begin{eqnarray}
\label{constr-J-a}
J_+ |\Psi^{(2s)}\rangle &=& 0\,, \\ [6pt]
\label{constr-J3a}
(J_3-s) |\Psi^{(2s)}\rangle &=& 0\,,
\end{eqnarray}
where the operators $J_+$ and $J_3$
are expressed via harmonic derivatives \p{D-u}, \p{D-w} in \p{Ja-ab3}.
Recall that the vector $|\Psi^{(2s)}\rangle$ also obeys the conditions \p{C4-constr}:
\begin{equation}
\label{C4-constr1}
\ell\,|\Psi^{(2s)}\rangle \ = \  \mu\,|\Psi^{(2s)}\rangle \,,\qquad \tilde\ell\,|\Psi^{(2s)}\rangle \ = \ \mu\,|\Psi^{(2s)}\rangle \, .
\end{equation}

Now, we show that the vectors of the states $|\Psi^{(2s)}\rangle$ are
realized as fields. In the representation \p{vw-dif} the
corresponding fields in the light-cone frame are the functions
$\Psi^{(2s)}(u^\pm, v^\pm)$ of four $\mathrm{SU}(2)$ spinors
$u_i^\pm$, $v_{\underline{i}}^\pm$.
Then, it is natural to present the solution of equations \p{C4-constr1} by using $\delta$-functions
\begin{equation}
\label{sol-0}
\Psi^{(2s)}(u^\pm, v^\pm)=\delta(\ell- \mu) \delta(\tilde\ell-\mu)\Phi^{(2s)}(u^\pm, v^\pm) \, ,
\end{equation}
where the arguments of the field $\Phi^{(2s)}(u^\pm, v^\pm)$ satisfy \p{harm-cond-1a}, \p{harm-cond-2b},
and it means that the field $\Psi^{(2s)}(u^\pm, v^\pm)$ is a function on the $\mathrm{SU}(2)\,{\otimes}\,\mathrm{SU}(2)$ group.

Using the relations \p{Ja-ab3}, one rewrites the remaining conditions \p{constr-J-a} and
\p{constr-J3a} in the form
\begin{eqnarray}
\label{constr-J-b}
\left( D^{++}_u +D^{++}_v\right) \Phi^{(2s)}(u^\pm, v^\pm) &=& 0\,, \\ [6pt]
\label{constr-J3b}
\left( D^{0}_u +D^{0}_v-2s\right) \Phi^{(2s)}(u^\pm, v^\pm) &=& 0\,.
\end{eqnarray}

Equation \p{constr-J3b} means the $\mathrm{U}(1)$ covariance of the
field $\Phi^{(2s)}(u^\pm,v^\pm)$:
\begin{equation}
\label{u1-field}
\Phi^{(2s)}(e^{\pm i\varphi}u^\pm, e^{\pm i\alpha}v^\pm)=e^{2si\alpha}\Phi^{(2s)}(u^\pm, v^\pm) \,.
\end{equation}
The charge $(2s)$ in the notation of the field $\Phi^{(2s)}$ reflects the property \p{u1-field} of this field.
Transformations of the arguments of the field $\Phi^{(2s)}(u^\pm,v^\pm)$ in \p{u1-field} are obtained from
the right action on the matrices \p{u-matrix} by the diagonal unitary matrix $h$:
\begin{equation}
\label{uv-u1}
u_i{}^J\,\to \,u_i{}^K h_K{}^J\,,\qquad v_{\underline{i}}{}^J\,\to \,v_{\underline{i}}{}^K h_K{}^J\,,\qquad\quad
h=\parallel\! h_K{}^J \!\parallel:=
\left(\!
\begin{array}{cc}
e^{ \,i\alpha} & 0 \\
0 & e^{- i\alpha} \\
\end{array}
\!\right)
,
\end{equation}
where $K,J=(+,-)$.
In the standard stereographic parametrization of the $SU(2)$ matrices
\begin{equation}
\label{su2-par}
\begin{array}{rcl}
\left(\!
\begin{array}{cc}
u_1^+ & u_1^- \\
u_2^+ & u_2^- \\
\end{array}
\!\right)&=&\displaystyle{\frac{1}{\sqrt{1+t_1\bar t_1}}}
\left(\!
\begin{array}{cc}
1 & -\bar t_1 \\
t_1 & 1 \\
\end{array}
\!\right)
\left(\!
\begin{array}{cc}
e^{ \,i(\psi+\varphi)} & 0 \\
0 & e^{- i(\psi+\varphi)} \\
\end{array}
\!\right)
, \\ [10pt]
\left(\!
\begin{array}{cc}
v_1^+ & v_1^- \\
v_2^+ & v_2^- \\
\end{array}
\!\right)&=&\displaystyle{\frac{1}{\sqrt{1+t_2\bar t_2}}}
\left(\!
\begin{array}{cc}
1 & -\bar t_2 \\
t_2 & 1 \\
\end{array}
\!\right)
\left(\!
\begin{array}{cc}
e^{ \,i(\psi-\varphi)} & 0 \\
0 & e^{- i(\psi-\varphi)} \\
\end{array}
\!\right)
,
\end{array}
\end{equation}
the transformation \p{uv-u1} is represented by the phase shift $\psi\to\psi+\alpha$.
Moreover, due to the fact that the field $\Phi^{(2s)}(u^\pm, v^\pm)$ has a fixed $\mathrm{U}(1)$-charge equal to $2s$,
its dependence on the phase variable $\psi$ is factorized:
\begin{equation}
\label{field-fact}
\Phi^{(2s)}(u^\pm, v^\pm)=e^{2si\psi}\hat\Phi(t_1,t_2,\bar t_1,\bar t_2,\varphi) \,.
\end{equation}
The field $\hat\Phi(t_1,t_2,\bar t_1,\bar t_2,\varphi)$ on the
right-hand side of equality \p{field-fact} is the function on the
coset $[\mathrm{SU}(2)\,{\otimes}\,\mathrm{SU}(2)]/\mathrm{U}(1)$
where the variable $\psi$ is the coordinate of the stability
subgroup $\mathrm{U}(1)$. Thus, the field $\Phi^{(2s)}(u^\pm,
v^\pm)$ having a fixed $\mathrm{U}(1)$-charge is in a one-to-one
correspondence with the function on the coset space
$[\mathrm{SU}(2)\,{\otimes}\,\mathrm{SU}(2)]/\mathrm{U}(1)$
\cite{GIKOS,GIOS}\footnote{Various aspects of functions on such a
coset are discussed in \cite{Kuz}.}. For this reason, we may refer
to the variables $u_i^\pm$, $v_{\underline{i}}^\pm$ used here as the
$[\mathrm{SU}(2)\,{\otimes}\,\mathrm{SU}(2)]/\mathrm{U}(1)$
harmonics. Since the variables $u_i^\pm$, $v_{\underline{i}}^\pm$
consist of twice the number of harmonics used in \cite{GIKOS,GIOS},
the space parameterized by these four $\mathrm{SU}(2)$ spinors can
be called the bi-harmonic space.

Note that a slightly different type of the bi-harmonic space was
previously used in the study of various supersymmetric models. For
example, two types of harmonics were employed in \cite{IvSut} for
constructing an off-shell superfield formulation of the $2D, (4,4)$
sigma-model. In those papers, the harmonics were used to parameterize
the coset space $\mathrm{SU}_L(2)/\mathrm{U}_L(1) \otimes
\mathrm{SU}(2)_R/\mathrm{U}_R(1)$, where the spaces
$\mathrm{SU}_L(2)/\mathrm{U}_L(1)$ and
$\mathrm{SU}(2)_R/\mathrm{U}_R( 1) $ were associated with the harmonics
$u^{\pm,0}$ and $v^{0,\pm}$, respectively, having the charges of
different $\mathrm{U}(1)$ groups. As a result, the fields on this
harmonic coset have two $\mathrm{U}(1)$ charges, which are defined
as eigenvalues of the operators $D^{0}_u$ and $D^{0}_v$. On the
other hand, in the case considered here, the field is defined by
equation \p{constr-J3b}, where the only $\mathrm{U}(1)$ charge $2s$
of the field $\Phi^{(2s)}(u^\pm, v^\pm)$ is given as the eigenvalue
of the $\mathrm{U}(1)$ generator $D^{0}=D^{0}_u +D^{0}_v$. Besides,
the $\mathrm{U}(1)$ charges $(\pm)$ of the two pairs of harmonics
$u^\pm$, $v^\pm$ coincide unlike the harmonics in \cite{IvSut}. As
discussed above, this means that the field $\Phi^{(2s)}(u^\pm,
v^\pm)$ of a special type is defined on bi-harmonic space where the
coordinates $u^\pm$ and $v^\pm$ parameterize the coset
$[\mathrm{SU}(2)\,{\otimes}\,\mathrm{SU}(2)]/\mathrm{U}(1)$, as
shown in \p{u1-field}. Another type of bi-harmonics was used, e.g. in
\cite{BII}, to describe the effective actions ${\cal N}=4$ SYM theory
(see the details in \cite{BII} and the references therein).

Now we describe the general solution to equations
\p{constr-J-b} and \p{constr-J3b}.

First, we note that any function of the variables
\begin{equation}
\label{y-+}
y_{i\underline{j}}:=u_{i}^+ v_{\underline{j}}^- -  u_{i}^- v_{\underline{j}}^+
\end{equation}
satisfies eqs. \p{constr-J-b}, \p{constr-J3b} for $s=0$.

Then, in general, the field $\Phi^{(2s)}(u^\pm, v^\pm)$,
obeying equations \p{constr-J-b}, \p{constr-J3b} for $2s\in
\mathbb{Z}{\geq}0$, is written in the form
\begin{equation}
\label{field-gen++}
\Phi^{(2s)}(u^\pm, v^\pm) \ = \ \sum_{r=0}^\infty
\Phi^{(2s)}_{k(r)\,\underline{l}(r)}(u^+, v^+) \,y^{k(r)\,\underline{l}(r)} \,,
\end{equation}
where
\begin{equation}
\label{field-gen++a}
\Phi^{(2s)}_{k(r)\,\underline{l}(r)}(u^+, v^+) \ = \ \sum_{\substack{p, q=0, \\ p+q=2s }}^{2s}
\phi_{k(r)\,\underline{l}(r)}^{i(p)\, \underline{j}(q)} u_{i(p)}^+ v_{\underline{j}(q)}^+  \,.
\end{equation}
Expressions \p{field-gen++} and \p{field-gen++a} use the following concise notation for the monomials:
\begin{equation}
\label{y-n}
u_{i(r)}^+:=u_{i_1}^+\ldots u_{i_r}^+ \,,\qquad
v_{\underline{i}(r)}^+:=v_{\underline{i}{}_1}^+\ldots v_{\underline{i}{}_r}^+ \,,\qquad
y^{i(r)\underline{j}(r)}:= y^{i_1\underline{j}{}_1}\ldots y^{i_r\underline{j}{}_r}\,,
\end{equation}
and we use the standard convention $y^{i\underline{j}}=\epsilon^{ik}\epsilon^{\underline{j}\underline{l}}y_{k\underline{l}}$ for raising and lowering the $\mathrm{SU}(2)$ indices.

The field $\Phi^{(2s)}(u^\pm,v^\pm)$ in \p{field-gen++}
that satisfies  \p{constr-J-b} and \p{constr-J3b}
is a linear combination with the constant coefficients $\phi_{k(r)\,\underline{l}(r)}^{i(p)\, \underline{j}(q)}$
of an infinite number of basis states
$u_{i(p)}^+ v_{\underline{j}(q)}^+y_{k(r)\,\underline{l}(r)}$.
The corresponding combinations of these basis vectors allow us to define the space for the irreducible infinite spin $\mathfrak{iso}(1,5)$ representation in the light-cone frame.

As a solution of the irreducibility condition \p{C6-Cas},
we chose one of the $2s{+}1$ possible vectors in the space of the $\mathfrak{su}(2)$ irreps with spin $s$,
namely, we took the higher weight vector $|\Psi^{(2s)}\rangle$ defined
by conditions \p{constr-J-a}, \p{constr-J3a}.
This choice does not lead to loss of generality.
The remaining $2s$ vectors are obtained from this vector $|\Psi^{(2s)}\rangle$
by acting of the operator $(J_-)^k$ at $k=1,\ldots,2s$.
In the representation \p{Ja-ab3}, \p{constr-J-b} and \p{constr-J3b} the fields $\Phi^{(2s-2k)}$
are obtained by the action of the operator $(D^{--})^k$ on the field $\Phi^{(2s)}$:
$\Phi^{(2s-2k)}=(D^{--})^k\Phi^{(2s)}$.
Note that the action of $(D^{--})^k$ decreases the degree of the polynomial $\Phi^{(2s)}_{k(r)\,\underline{l}(r)}(u^+, v^+)$ \p{field-gen++a}
in the variables $(u^+,v^+)$ and increases it in the variables $(u^-,v^-)$.
Choosing any other fields $\Phi^{(2s-2k)}$, $k=1,\ldots,2s$ leads to the equivalent
infinite spin representations of $\mathfrak{iso}(1,5)$.

\setcounter{equation}0
\section{Field theory in the light-front coordinates}

In the previous section, we have developed a description of the irreducible $6D$ infinite spin
representations in the light-cone frame and shown that this description is formulated in terms
of fields in bi-harmonic space. Now we extend this analysis to the light-front coordinate system and construct the
corresponding field theory.

The formulation of the field theory on the light-front was proposed
by Dirac \cite{Dirac49}, its further development and applications
were considered by many authors (see e.g.
\cite{Bengtsson2Brink,Bengtsson2Linden,Metsaev05,PoSk} and the
references therein). The light-front is defined as the surface
$x^+=const$ in the six-dimensional Minkowski space $\mathbb{R}^{1,5}$.
It means that the coordinate $x^+$ is interpreted as a "time"
evolution parameter. Therefore, the role of the Hamiltonian in the case
under consideration is played by the operator
\begin{equation}
\label{P-H}
H=P^-\,.
\end{equation}

To define an infinite spin field in the light-front coordinates, we
will use the results of the previous section, where the
corresponding field is given by \p{field-gen++}, \p{field-gen++a}.
Note that the light-cone coordinate system is obtained from the
light-front coordinate system by vanishing the coordinates $x^{\hat
a}$ and fixing the coordinates $x^\pm$. Therefore, it is natural to
assume that the infinite spin field in the light-front coordinates
should have the form \p{field-gen++}, \p{field-gen++a}, where,
however, the coefficients $\phi_{k(r)\,\underline{l}(r)}^{i(p)
\underline{i}(q)}$ are functions of $x^\pm$ and $x^{\hat a}$. As a
result, the irreducible infinite spin field depending on the light-front
coordinates is defined as
\begin{equation}
\label{field-gen++1}
\Phi^{(2s)}(x^\pm,x^{\hat a},u^\pm, v^\pm)\ = \ \sum_{\substack{p, q=0, \\ p+q=2s }}^{2s} \sum_{r=0}^\infty
\phi_{k(r)\,\underline{l}(r)}^{i(p) \underline{i}(q)}(x^\pm,x^{\hat a}) \, u_{i(p)}^+ v_{\underline{j}(q)}^+ \,y^{k(r)\,\underline{l}(r)}\,.
\end{equation}
Taking into account a general principle of the light-cone dynamics
\cite{Dirac49}, one concludes that the equation of motion for the
field \p{field-gen++1} is the Schr\"{o}dinger-type equation
\begin{equation}
\label{Schr-eq}
\left(-i\frac{\partial}{\partial x^+}-H \right)\Phi^{(2s)}(x^\pm,x^{\hat a},u^\pm, v^\pm)=0 \,,
\end{equation}
where the coordinate $x^+$ plays the role of time.

As usual, the generators $P_{a}$ and $M_{ab}$ of $\mathfrak{iso}(1,5)$ in
the the light-front formulation are divided into kinematic and dynamic generators. One can show that
these divisions in the field realization \p{field-gen++1} have the form
\begin{itemize}
\item
Kinematic generators
\begin{equation}
\label{P-kin}
P^+=p^+\,,\qquad P^{\hat a}=p^{\hat a} \,,
\end{equation}
\begin{equation}
\label{M-kin}
M^{\hat a \hat b}=x^{\hat b}p^{\hat a} -x^{\hat a}p^{\hat b}+ S^{\hat a \hat b}\,,\qquad
M^{+\hat a}=x^{\hat a}p^+ + S^{+\hat a}\,,\qquad
M^{+-}=x^{-}p^+ + S^{+-} \,;
\end{equation}
\item
Dynamic generators
\begin{equation}
\label{P-dyn}
P^-=\frac{p^{\hat a}p^{\hat a}}{2p^+}=H\,,
\end{equation}
\begin{equation}
\label{M-dyn}
M^{-\hat a}=x^{\hat a}H -x^{-}p^{\hat a} + S^{-\hat a}\,,
\end{equation}
\end{itemize}
where
\begin{equation}
\label{P-dif}
p^{\hat a}=i\frac{\partial}{\partial x^{\hat a}}\,,\qquad p^{+}=-i\frac{\partial}{\partial x^{-}}
\end{equation}
and all spin parts of the Lorentz rotation generators  $S^{ab}=(S^{\hat a\hat b},S^{\pm\hat a},S^{+-})$
depend on the spinors  $u_i^\pm$ and $v_{\underline{i}}^\pm$ in the same way as the operators
\p{Mpm-eta-summ}, \p{Si-pm-ch}, \p{Si-pm-ch-1}, \p{Si-pm-ch-2}
depend on the spinors  $\xi_i^I$ and $\rho_{\underline{i}}^I$
\begin{eqnarray}
\label{Mab-lc}
S^{\hat a\hat b}&=& \frac{1}{2}\,\eta^{\mathrm{i}}_{\hat a\hat b}
\left[\frac{\partial}{\partial u_i^+} (\tau_{\mathrm{\,i}})_i{}^j u_j^+
+ \frac{\partial}{\partial u_i^-} (\tau_{\mathrm{\,i}})_i{}^j u_j^- \right]
+\frac{1}{2}\,\bar\eta^{\mathrm{i}^\prime}_{\hat a\hat b}
\left[\frac{\partial}{\partial v_{\underline{i}}^+} (\tau_{\mathrm{\,i}^\prime})_{\underline{i}}{}^{\underline{j}}v_{\underline{j}}^+
+\frac{\partial}{\partial v_{\underline{i}}^-} (\tau_{\mathrm{\,i}^\prime})_{\underline{i}}{}^{\underline{j}}v_{\underline{j}}^-\right] ,
\\ [5pt]
\label{M+a-lc}
S^{+\hat a}&=&
\frac{i}{\sqrt{2}}\left[
\frac{\partial}{\partial u_i^+}(\tau_{\hat a})_{i}{}^{\underline{j}}\epsilon_{\underline{j}\underline{k}}\frac{\partial}{\partial v_{\underline{k}}^-} -\frac{\partial}{\partial u_i^-}(\tau_{\hat a})_{i}{}^{\underline{j}}\epsilon_{\underline{j}\underline{k}}\frac{\partial}{\partial v_{\underline{k}}^+}\right],\ \ \ \mbox{at} \ \ \hat a=1,2,3\,,
\\ [5pt]
\label{M-a-lc}
S^{-\hat a}&=&
\frac{i}{\sqrt{2}}\left[v_{\underline{k}}^-\epsilon^{\underline{k}\underline{i}}(\tau_{\hat a})_{\underline{i}}{}^{j}u_{j}^+
-v_{\underline{k}}^+\epsilon^{\underline{k}\underline{i}}(\tau_{\hat a})_{\underline{i}}{}^{j}u_{j}^-\right]
,\ \ \ \mbox{at} \ \ \hat a=1,2,3\,,
\\ [5pt]
\label{M+4-lc}
S^{+4}&=&
\frac{1}{\sqrt{2}}\left[
\frac{\partial}{\partial u_i^+}\epsilon_{i\underline{k}}\frac{\partial}{\partial v_{\underline{k}}^-} -\frac{\partial}{\partial u_i^-}\epsilon_{i\underline{k}}\frac{\partial}{\partial v_{\underline{k}}^+}\right],
\\ [5pt]
\label{M-4-lc}
S^{-4}&=&
\frac{1}{\sqrt{2}}\left[v_{\underline{k}}^-\epsilon^{\underline{k}i}u_{i}^+
-v_{\underline{k}}^+\epsilon^{\underline{k}i}u_{i}^-\right],\qquad
\\ [5pt]
\label{M+--lc}
S^{+-}&=&
\frac{1}{2}
\left[u_i^+ \frac{\partial}{\partial u_i^+}
+ u_i^- \frac{\partial}{\partial u_i^-} +
v_{\underline{i}}^+ \frac{\partial}{\partial v_{\underline{i}}^+}
+v_{\underline{i}}^- \frac{\partial}{\partial v_{\underline{i}}^-} +4\right].
\end{eqnarray}
Turn attention that all the generators $S^{ab}=(S^{\hat a\hat b},S^{\pm\hat a},S^{+-})$
defined in \p{Mab-lc}-\p{M+--lc} have zero  $\mathrm{U}(1)$-charge.

After acting by the operator $p^+$ on equation \p{Schr-eq}, this equation takes the form\footnote{For a discussion of the operator $p^+$ invertibility,  see e.g. \cite{Sieg88,Sieg99}.}
\begin{equation}
\label{Dal-eq}
\Box\,\Phi^{(2s)}(x^\pm,x^{\hat a},u^\pm, v^\pm)=0 \,,
\end{equation}
where $\Box$ is the d'Alambertian operator in the six-dimensional Minkowski space in the light-front coordinates
\begin{equation}
\label{Dal-op}
\Box \ := \ 2\frac{\partial}{\partial x^+}\frac{\partial}{\partial x^-}-\frac{\partial}{\partial x^{\hat a}}\frac{\partial}{\partial x^{\hat a}} \,.
\end{equation}
Equation \p{Dal-eq} is the equation of motion corresponding to the action
\begin{equation}
\label{act-lc}
S \ = \ \int d^{\,6}x\, du\, dv\ \bar\Phi^{(-2s)}\Box\,\Phi^{(2s)} \,,
\end{equation}
where $d^{\,6}x = dx^+ dx^- d^{\,4}x$ is the $6D$ Minkowski space measure and  $du dv$
is the bi-harmonic space measure \cite{GIKOS,GIOS}. The function $\bar\Phi^{(-2s)}$ is obtained by complex conjugation of
the function $\Phi^{(2s)}$:
\begin{equation}
\label{cc-field}
\bar\Phi^{(-2s)} \ = \ (\Phi^{(2s)})^* \,.
\end{equation}
Integration over harmonics is defined by simple rules (see \cite{GIKOS,GIOS} for details).
The integral is is a linear operation and it does not vanish only for $\mathrm{SU}(2)$-scalars with
the following normalization condition:
\begin{equation}
\label{int-harm-0}
\int du = 1\,,\qquad  \int dv = 1\,.
\end{equation}
For all other harmonic monomials, the harmonic integral is equal to zero:
\begin{equation}
\label{int-harm-1}
\int du \, u^+_{(i_1}\ldots u^+_{i_m}u^-_{j_1}\ldots u^-_{j_n)}= 0\,,\qquad
\int dv \, v^+_{(\underline{i}_1}\ldots v^+_{\underline{i}_m}v^-_{\underline{j}_1}\ldots v^-_{\underline{j}_n)}= 0\,,
\end{equation}
at arbitrary integers $m$ and $n$ which are not equal to zero simultaneously.

Reality conditions \p{u-real} are now written as follows:
\begin{equation}
\label{harm-conj}
(u^{i\pm})^*=\pm\, u_i^\mp\,,\qquad  (v^{i\pm})^*=\pm\, v_i^\mp\,.
\end{equation}
Therefore, at complex conjugation the charge $2s$ of the harmonic field $\Phi^{(2s)}$ changes to
$-2s$ in accordance with \p{cc-field}. As a result, the integrand in \p{act-lc} has a zero harmonic charge
as it should be for the non-vanishing harmonic integral.\footnote{Note that in the harmonic superspace approach to
${\cal N}=2$ supersymmetric field theories \cite{GIKOS,GIOS} another rule of conjugation was used that
combines complex conjugation with an antipodal map. However, in the case under consideration, the
ordinary complex conjugation is totally appropriate.}

In expansion of the harmonic field \p{field-gen++1} the indices $i$ and $\underline{i}$ of the
component fields
$\phi^{(i(m)j(n))(\underline{i}(k)\underline{j}(n))}(x)$ are half of the $6D$
$\mathrm{SU}^*(4)$-indices $\alpha$. It means, for half-integer $s$, the harmonic field
$\Phi^{(2s)}(x,u^\pm,v^\pm)$ is an odd order polynomial in $u^\pm$, $v^\pm$  and describes half-integer spin fields with an odd number of indices.
Therefore, the fermionic fields should be endowed by the corresponding odd statistics. Besides, in the
fermionic case, the natural Lagrangian is the one of the first order in space derivatives. This type of Lagrangian in
the light-front formalism is obtained from the Lagrangian \p{act-lc} by replacement
\begin{equation}
\label{Psi-Phi}
\Psi^{(2s)} \ = \ \sqrt{p^+}\,\Phi^{(2s)} \,.
\end{equation}
Then, for the field $\Psi^{(2s)}$ expression \p{act-lc} leads to the following Lagrangian:\footnote{For discussion of the light-front describing the fields with different statistics see e.g.
\cite{Sieg88,Sieg99}.}
\begin{equation}
\label{La-lc}
\bar{\Psi}^{(-2s)}\left(p^- - H \right) \Psi^{(2s)} \,.
\end{equation}

Thus, the action \p{act-lc} determines the field dynamics of
infinite spin fields on the light front. A specific feature of the
obtained theory is its formulation in terms of harmonic variables.

\setcounter{equation}0
\section{Summary}

We have developed the $6D$ Minkowski space infinite spin free
Lagrangian field theory in the light-cone formalism. First, we have
studied this theory in the light-cone frame and unexpectedly found
that the corresponding infinite spin field is a function on a
special bi-harmonic space associated with the coset
$[\mathrm{SU}(2)\,{\otimes}\,\mathrm{SU}(2)]/\mathrm{U}(1)$. Second,
the result obtained was generalized to the light-front coordinate
system, where the infinite spin field is described by the function
$\Phi^{(2s)}(x^\pm,x^{\hat a},u^\pm, v^\pm)$ \p{field-gen++1}
depending on the light-front coordinates and harmonics. Representations
of all the $6D$ Poincar\'e group generators in this coordinate
system are constructed. The field equation of motion in the light-front
coordinate system has the form of Schr\"{o}dinger-type equation
\p{Schr-eq} with the Hamiltonian \p{P-H}. The corresponding action
is given by \p{act-lc}.

The harmonic light-front approach formulated in this paper opens a
possibility to construct an interacting theory for $6D$ infinite
spin fields. One can expect that introducing an interaction will
lead to a modification of the dynamic generators \p{P-dyn}, \p{M-dyn}
by the interaction terms (see the description of interactions in
the light-front formalism, e.g., in \cite{PoSk} and the references
therein). In particular, the Hamiltonian \p{P-dyn} should go to
\begin{equation}
\label{P-P-int}
H\quad \rightarrow \quad H+H_{\mathrm{int}}\,.
\end{equation}
The harmonic formalism allows one from the very beginning to make some
simple predictions on the structure of the interacting Hamiltonian
$H_{\mathrm{int}}$. To preserve zero harmonic charge of the action,
this Hamiltonian should have zero harmonic charge as well. It
immediately means that an arbitrary order self-interaction of the same harmonic fields
$\Phi^{(2s)}$ is possible only for $s=0$ if other charged
harmonic quantities in the action are absent.
Self-interaction of charged fields $\Phi^{(2s)}$, $s\neq0$ can only be of an even order, such as
$\,\sim \bar{\Phi}^{(-2s)}\bar{\Phi}^{(-2s)}{\Phi}^{(2s)}{\Phi}^{(2s)}$.
Although for fields with different charges there is an additional choice in the structure of the interaction Lagrangian.
For example, the following interacting terms
$\,\sim \bar{\Phi}_1^{(-2s)}\left({\Phi}_2^{(0)}+\bar{\Phi}_2^{(0)}\right){\Phi}_1^{(2s)}$
or
$\,\sim\!\left({\Phi}_1^{(q_1)}{\Phi}_2^{(q_2)}{\Phi}_3^{(q_3)}+c.c.\right)$ at $q_1+q_2+q_3=0$
are
allowed in the action. In general, the requirement of zero charge of
interacting contributions to the action controls both charges of
interacting fields and their number. We plan to construct
interacting infinite spin $6D$ theories in the forthcoming works.

We also think that the appearance of bi-harmonic space in the
infinite spin representations of $\mathfrak{iso}(1,5)$ can indicate
the existence of manifest $\mathcal{N}{=}\,(1,0)$
supersymmetrization of the theory \p{act-lc},
\p{P-P-int}.\footnote{See, e.g., harmonic superfield formulation of the
six-dimensional $\mathcal{N}{=}\,(1,0)$ and $\mathcal{N}{=}\,(1,1)$
supersymmetric theories in \cite{BIS} and references therein.}

\section*{Acknowledgements}
The authors are grateful to E.A.\,Ivanov for discussing the aspects
of the harmonic formalism. The work of ILB and SAF is supported by
the Russian Science Foundation, project No 21-12-00129. The work of API was
partially supported by the Ministry of Education of the Russian
Federation, project FEWF-2020-0003.

\section*{Appendix A. Calculation of the Casimir operator $C_6$}
\def\theequation{A.\arabic{equation}}
\setcounter{equation}0

\quad\,

The six-order Casimir operator in the $6D$ massless theory is given by \p{C6-Casimir}. The derivation of this
operator in \cite{BFIP} was based on the relations:
\begin{equation}
\label{C6-Casimir-def}
C_6=\frac{1}{64}\, \Upsilon^a\Upsilon_a\,,\qquad
\Upsilon_a=\varepsilon_{abcdef}P^{b}  M^{cd} M^{ef}
\end{equation}
These expressions show that only the spin part $S_{ab}$ \p{M-1} of the Lorentz group generators
$M_{ab}$ \p{M-gen} contributes to the Casimir operator \p{C6-Casimir-def}.
However, if we substitute $S_{ab}$ for $M_{ab}$ into \p{C6-Casimir},
an incorrect result is obtained, since when passing from \p{C6-Casimir-def} to \p{C6-Casimir},
one has to rearrange the operators $ P_{c}$ and $M_{ab} $ using commutators.
At the replacement $M_{ab}\to S_{ab}$, a correct expression $C_6$ can be obtained only after preliminary "untangling" of the generators
$P_{c}$ и $M_{ab}$ in expression \p{C6-Casimir}.

We will act in the following way. First, we rearrange with the help
of commutation relations all the operators $P_{c}$ to the right on all the operators $M_{ab}$ in
expression \p{C6-Casimir}. Second, after such an ordering is done, we replace the operator $M_{ab}$ by
the operator $S_{ab}$ in the obtained expression.

Using the commutator $[M_{ab},\Pi_{c}]=i\left(\eta_{ac}\Pi_{b}-\eta_{bc}\Pi_{a}\right)$, we rearrange
the operators $\Pi_{a}$ to the right in the first term of expression \p{C6-Casimir}. All the terms
proportional to $2\Pi_{[a} \Pi_{b]}=[\Pi_a, \Pi_b] = -i \, M_{ab} \, P^2 $ can be omitted for a
massless representation where $P^2=0$. Besides, since we consider irreducible infinite spin
for which the condition \p{C4-const} holds,
we replace the operator $C_4$ by its eigenvalue $-\mu^2$ in the second term of \p{C6-Casimir}.
Now all operators $M_{ab}$ to the left of the operators $\Pi_a$ are replaced by the operators $S_{ab}$.
As a result, one obtains
\begin{equation}
\label{C6-Casimir-1}
C_6=
-\, S^{(b}{}_{a}\, S^{c)a}\Pi_b \Pi_c
\ - \ \frac12\,\mu^2\,\Big(S^{bc}S_{bc}-8\Big) \,.
\end{equation}

Using the relations for the $\sigma$-matrices from \cite{BFI-21}, the identity
\begin{equation}
\label{sigma-id}
(\tilde\sigma^b{}_a)^\alpha{}_\beta (\tilde\sigma^{ca})^\gamma{}_\delta
= \frac14\,\eta^{bc}\left(\delta^\alpha_\beta \delta^\gamma_\delta -2 \delta^\alpha_\delta \delta^\gamma_\beta\right)
+\frac12 (\tilde\sigma^{(b})^{\alpha\gamma} (\sigma^{c)})_{\beta\delta}
+\frac12\,\delta^\alpha_\delta (\tilde\sigma^{bc})^\gamma{}_\beta
- \frac12\,\delta^\gamma_\beta (\tilde\sigma^{bc})^\alpha{}_\delta\,,
\end{equation}
the commutation relations \p{com-sp} and the realization \p{M-1} for the operators $S_{ab}$, one gets the equality
\begin{equation}
\label{S2-exp1}
S^{(b}{}_aS^{\,c)a}
= \eta^{bc}\left[\frac12\,(\xi_{(I}\rho_{K)})(\xi^{I}\rho^{K})
+i (\xi^{I}\rho_I)\right]
+\frac12\,\xi_\alpha^I \rho^\beta_I \xi_\gamma^J \rho^\delta_J
(\tilde\sigma^{(b})^{\alpha\gamma}(\sigma^{c)})_{\beta\delta}\,,
\end{equation}
which leads to
\begin{equation}
\label{S2-exp2}
S^{bc}S_{bc}
= - \frac12\,(\xi^{I}\rho_I)^2
+4i (\xi^{I}\rho_I)
+2 (\xi_{(I}\rho_{K)})(\xi^{I}\rho^{K})\,,
\end{equation}
where the notation $(\xi^{I}\rho_{K}):=\xi_\alpha^{I}\rho^\alpha_{K}$ has been used.
After substituting  \p{S2-exp1} and  \p{S2-exp2} into \p{C6-Casimir-1}, one gets
\begin{equation}
\label{C6-Casimir-2}
C_6=
\mu^2\left[\frac14\,(\xi^{I}\rho_I)^2-\frac12\,(\xi_{(I}\rho_{J)})(\xi^{I}\rho^{J})
-i (\xi^{I}\rho_I)+4\right]
-\frac12\,\xi_\alpha^I \rho^\beta_I \xi_\gamma^J \rho^\delta_J
(\tilde\sigma^{b})^{\alpha\gamma}(\sigma^{c})_{\beta\delta}\Pi_b \Pi_c \,.
\end{equation}
The last term in this expression is represented in the following form:
\begin{equation}
\label{2-C6-Casimir-2}
-\frac12\,\xi_\alpha^I \rho^\beta_I \xi_\gamma^J \rho^\delta_J
(\tilde\sigma^{b})^{\alpha\gamma}(\sigma^{c})_{\beta\delta}\Pi_b \Pi_c =
-\frac{i}{2}\,\mu^2 (\xi^{I}\rho_I)
+\frac14\, (\xi^I\tilde\sigma^b\xi_I)(\rho_J\sigma^c\rho^J)\Pi_b \Pi_c\,,
\end{equation}
where
$(\xi^I\tilde\sigma^b\xi_I):=\xi_\alpha^I(\tilde\sigma^b)^{\alpha\beta}\xi_{\beta I}$,
$(\rho_J\sigma^c\rho^J):=\rho^\alpha_J(\sigma^c)_{\alpha\beta}\rho^{\beta J}$.

The last step in deriving the expression for $C_6$ is to move the operators $P_m$ to the right in expression \p{2-C6-Casimir-2}.
Using the equality $\Pi_a = M_{ba}\, P^{b} - 5 i \, P_a$, we write the expression  $\Pi_b \Pi_c$ in the form:
\begin{equation}
\label{Pi-2}
\Pi_b \Pi_c=M_{eb}M_{fc}P^eP^f-6iM_{eb}P^eP_c -5iM_{ec}P^eP_b-30P_bP_c\,.
\end{equation}
Now that all operators $M_{ab}$ on the right side of \p{Pi-2} are to the left of all operators $P_c$, we replace the operators $M_{ab}$ with their spin parts $S_{ab}$.

After such a replacement $M_{ab}\to S_{ab}$, where $S_{ab}$ are defined in \p{M-1},  the substitution \p{Pi-2} into \p{2-C6-Casimir-2} and \p{C6-Casimir-2} and using the equalities
\begin{eqnarray}
\label{c-a}
\frac{1}{4}\,(\xi^I\tilde\sigma^b\xi_I)(\rho_J\sigma^c\rho^J)S_{eb}S_{fc}P^e P^f &=&
\mu^2\left[-\frac14\,(\xi^{I}\rho_I)^2+2i (\xi^{I}\rho_I)-6\right], \\ [6pt]
\label{c-b}
-\frac{3i}{2}\,(\xi^I\tilde\sigma^b\xi_I)(\rho_J\sigma^c\rho^J)S_{eb}P^e P_c &=&
\mu^2\left[-3i (\xi^{I}\rho_I)+12\right], \\ [6pt]
\label{c-c}
-\frac{5i}{4}\,(\xi^I\tilde\sigma^b\xi_I)(\rho_J\sigma^c\rho^J)S_{ec}P^e P_b &=&
\mu^2\left[\frac{5i}{2} (\xi^{I}\rho_I)+20\right], \\ [6pt]
\label{c-d}
-\frac{15}{2}\,(\xi^I\tilde\sigma^b\xi_I)(\rho_J\sigma^c\rho^J) P_b P_c &=&  -30\mu^2\,,
\end{eqnarray}
which are valid at $P^2=0$ and $(\xi^I\tilde\sigma^b\xi_I)(\rho_J\sigma^c\rho^J)P_b P_c=4\mu^2$
(second equality is the condition \p{C4-const} for the fourth-order Casimir operator), one obtains
\begin{equation}
\label{C6-Casimir-fin}
C_6= -\frac12\,\mu^2(\xi_{(I}\rho_{J)})(\xi^{I}\rho^{J}) \,.
\end{equation}
When using the operators \p{Ta-def}, this final expression \p{C6-Casimir-fin} is represented as
\begin{equation}
\label{C6-Casimir-fin1}
C_6= -\mu^2J_{\mathrm{i}}J_{\mathrm{i}}\,,
\end{equation}
which is the same as \p{C6-Cas}.

\section*{Appendix B. Spinor part of the $\mathfrak{so}(1,5)$-generators}
\def\theequation{B.\arabic{equation}}
\setcounter{equation}0

\quad\,

We consider a representation where the $(4{\times}4)$ $\sigma$-matrices \cite{KuTow,IR,BFI-21}
$\sigma^a=\| (\sigma^a)_{\alpha\dot\beta}\|$,
$\tilde\sigma^a=\| (\tilde\sigma^a)^{\dot\alpha\beta}\|$ with the $6D$ vector index $a=0,1,\ldots,5$
and spinor indices  $\alpha,\dot\alpha=1\dots,4$ are realized in the form of the following matrices:
\begin{equation}
\label{sigma-a}
\sigma^a=(\sigma^0,\sigma^{\hat a},\sigma^5)\,,\qquad \tilde\sigma^a=(\sigma^0,-\sigma^{\hat a},-\sigma^5)\,,\qquad \hat a=1,\ldots,4\,,
\end{equation}
where
\begin{equation}
\label{sigma-a-r}
\sigma^0=1_4\,, \qquad \sigma^{\hat a}=\tau_2\otimes \tau_{\hat a} \ \ \mbox{at} \ \ \hat a=1,2,3\,,
\qquad \sigma^4=-\tau_1\otimes 1_2\,,\qquad
\sigma^5=\tau_3\otimes 1_2
\end{equation}
and $\tau_{1,2,3}$ are the Pauli matrices.\footnote{ In \cite{KuTow,BFI-21} the
representation $\sigma^1=\tau_1\otimes 1_2$, $\sigma^{\hat a}=
\tau_2\otimes \tau_{\hat a-1}$ at $\hat a=2,3,4$ and $\sigma^0$, $\sigma^5$ was used as in \p{sigma-a-r}.
That is, distinction between the representations \cite{KuTow,BFI-21} and  \p{sigma-a-r} lies in the difference
in the notation of the four space coordinates labeled by $\hat a$. However, the representation
\p{sigma-a-r} is more convenient when using the standard realizations \p{eta} for the 't\,Hooft symbols.}
The antisymmetric $\sigma$-matrices with non-dotted spinor subscripts and superscripts are defined as follows:
\begin{equation}
\label{sigma-m}
(\sigma^{a})_{\alpha\beta}=(\sigma^{a})_{\alpha\dot\gamma}(B^{-1})_\beta{}^{\dot\gamma}\,,\qquad
(\tilde\sigma^{a})^{\alpha\beta}=B_{\dot\gamma}{}^\alpha(\tilde\sigma^{a})^{\dot\gamma\beta}\,,
\end{equation}
where
$B=\| B_{\dot\alpha}{}^{\beta}\|=1_2\otimes i\tau_{2}=
\left(
\begin{array}{cc}
i\tau_2 & 0 \\
0 & i\tau_2 \\
\end{array}
\right)
$
is the matrix defining complex conjugation of the $6D$ Weyl spinors \cite{KuTow,IR,BFI-21}. In particular,  the
matrices $\sigma^\pm=(\sigma^0\pm\sigma^5)/\sqrt{2}$,
$\tilde\sigma^\pm=(\tilde\sigma^0\pm\tilde\sigma^5)/\sqrt{2}$ with undotted indices  have the form:
\begin{equation}
\label{sigma+-}
(\sigma^{+})_{\alpha\beta}=
\sqrt{2}\left(\!
\begin{array}{cc}
i\tau_2 & 0 \\
0 & 0 \\
\end{array}
\!\right)=
\left(\!
\begin{array}{cc}
\sqrt{2}\epsilon_{ij} & 0 \\
0 & 0 \\
\end{array}
\!\right),\quad
(\sigma^{-})_{\alpha\beta}=
\sqrt{2}\left(\!
\begin{array}{cc}
0 & 0 \\
0 & i\tau_2 \\
\end{array}
\!\right)=
\left(\!
\begin{array}{cc}
0 & 0 \\
0 & \sqrt{2}\epsilon_{\underline{i}\underline{j}} \\
\end{array}
\!\right),
\end{equation}
\begin{equation}
\label{t-sigma+-}
(\tilde\sigma^{+})^{\alpha\beta}=
\sqrt{2}\left(\!
\begin{array}{cc}
0 & 0 \\
0 & -i\tau_2 \\
\end{array}
\!\right)=
\left(\!
\begin{array}{cc}
0 & 0 \\
0 & \sqrt{2}\epsilon^{\underline{i}\underline{j}} \\
\end{array}
\!\right),\quad
(\tilde\sigma^{-})^{\alpha\beta}=
\sqrt{2}\left(\!
\begin{array}{cc}
-i\tau_2 & 0 \\
0 & 0 \\
\end{array}
\!\right)=
\left(\!
\begin{array}{cc}
\sqrt{2}\epsilon^{ij} & 0 \\
0 & 0 \\
\end{array}
\!\right).
\end{equation}
Also we use the standard representation for the 't\,Hooft symbols
$\eta^{\mathrm{i}}_{\hat a\hat b}=-\eta^{\mathrm{i}}_{\hat b\hat a}$,
$\mathrm{i}=1,2,3$ and $\bar\eta^{\mathrm{i}^\prime}_{\hat a\hat b}=
-\bar\eta^{\mathrm{i}^\prime}_{\hat b\hat a}$, $\mathrm{i}^\prime=1,2,3$
(see e.g. \cite{tHooft,IR1,BFIP})
\begin{equation}
\label{eta}
\eta^{\mathrm{i}}_{\hat a\hat b}=
\left\{
\begin{array}{l}
\epsilon_{\mathrm{i}\hat a\hat b} \qquad \hat a,\hat b=1,2,3,\\
\,\delta_{\mathrm{i}\hat a} \qquad \,\hat b=4,
\end{array}
\right.
\qquad\quad
\bar\eta^{\mathrm{i}^\prime}_{\hat a\hat b}=
\left\{
\begin{array}{l}
\,\,\epsilon_{\mathrm{i}^\prime \hat a\hat b} \qquad \hat a,\hat b=1,2,3,\\
-\delta_{\mathrm{i}^\prime \hat a} \qquad \hat b=4.
\end{array}
\right.
\end{equation}

First, we will consider the $\mathfrak{so}(4)$-part of the generators \p{M-1}, i.e. the operators
\begin{equation}
\label{M-1-a}
S_{\hat a\hat b} = \xi_\alpha^I (\tilde\sigma_{\hat a\hat b})^{\alpha}{}_{\beta}\rho^\beta_I\,,\ \ \ {\hat a}=1,2,3,4\,.
\end{equation}
These six generators  $S_{\hat a\hat b}$ are written as the sum
\begin{equation}
\label{M-Mpm}
S_{\hat a\hat b}=S_{\hat a\hat b}^{(+)} + S_{\hat a\hat b}^{(-)}\,,
\end{equation}
where the $\mathrm{SO}(4)$-(anti-)self-dual parts
$
S^{(\pm)}_{\hat a\hat b} = \pm{\displaystyle \frac12}\,\epsilon_{\hat a\hat b\hat c\hat d}S^{(\pm)}_{\hat c\hat d}
$
are expressed in terms of the $\mathrm{SO}(3)$-vectors $S^{(+)}_{\mathrm{\,i}}$, $S^{(-)}_{\mathrm{\,i}^\prime}$:
\begin{equation}
\label{Mpm-eta}
S^{(+)}_{\hat a\hat b} = -\eta^{\mathrm{i}}_{\hat a\hat b}S^{(+)}_{\mathrm{\,i}}\,,\qquad
S^{(-)}_{\hat a\hat b} = -\bar\eta^{\mathrm{i}^\prime}_{\hat a\hat b}S^{(-)}_{\mathrm{\,i}^\prime}\,,
\end{equation}
if we use the 't\,Hooft symbols \p{eta}. Thus, the generator \p{M-1-a} has the expansion
\begin{equation}
\label{Mpm-eta-summ}
S_{\hat a\hat b}= -\eta^{\mathrm{i}}_{\hat a\hat b}S^{(+)}_{\mathrm{\,i}}
-\bar\eta^{\mathrm{i}^\prime}_{\hat a\hat b}S^{(-)}_{\mathrm{\,i}^\prime}\,,
\end{equation}
where the operators $S^{(+)}_{\mathrm{\,i}}$ и $S^{(-)}_{\mathrm{\,i}^\prime}$ form two $\mathfrak{su}(2)$ algebras:
\begin{equation}
\label{M-M-2}
[S^{(+)}_{\mathrm{\,i}},S^{(+)}_{\mathrm{\,j}}]=i\epsilon_{\mathrm{i}\mathrm{j}\mathrm{k}}S^{(+)}_{\mathrm{\,k}}\,,\qquad
[S^{(-)}_{\mathrm{\,i}^\prime},S^{(-)}_{\mathrm{\,j}^\prime}]=
i\epsilon_{\mathrm{i}^\prime\mathrm{j}^\prime\mathrm{k}^\prime}S^{(-)}_{\mathrm{\,k}^\prime}\,,\qquad
[S^{(+)}_{\mathrm{\,i}},S^{(-)}_{\mathrm{\,j}^\prime}]=0\,.
\end{equation}

Using the equalities
$\eta^{\mathrm{i}}_{ab}\eta^{\mathrm{j}}_{ab}=4\delta^{\mathrm{i}\mathrm{j}}$,
$\bar\eta^{\mathrm{i}^\prime}_{ab}\bar\eta^{\mathrm{j}^\prime}_{ab}=
4\delta^{\mathrm{i}^\prime\mathrm{j}^\prime}$ and
$\eta^{\mathrm{i}}_{ab}\bar\eta^{\mathrm{j}^\prime}_{ab}=0$,
one finds the inverse to \p{Mpm-eta} relations
\begin{equation}
\label{Mpm-eta-inv}
S^{(+)}_{\mathrm{\,i}} = -\frac14\,\eta^{\mathrm{i}}_{\hat a\hat b}S_{\hat a\hat b}\,,\qquad
S^{(-)}_{\mathrm{\,i}^\prime} = -\frac14\,\bar\eta^{\mathrm{i}^\prime}_{\hat a\hat b}S_{\hat a\hat b}\,.
\end{equation}
However, using \p{sigma-a}, \p{sigma-a-r}, \p{sigma-m} and \p{eta} we obtain that the matrices present in the definition of the generators \p{Mpm-eta-inv} have only one diagonal $2{\times}2$ matrix block:
\begin{equation}
\label{eta-sigma1}
-\frac14\,\eta^{\mathrm{i}}_{\hat a\hat b}(\tilde\sigma_{\hat a\hat b})^{\alpha}{}_{\beta}=
\frac{i}{2}\left(
\begin{array}{cc}
-\tau^T_{\mathrm{\,i}} & 0 \\
0 & 0 \\
\end{array}
\right)
\,,\qquad
-\frac14\,\bar\eta^{\mathrm{i}^\prime}_{\hat a\hat b}(\tilde\sigma_{\hat a\hat b})^{\alpha}{}_{\beta}=
\frac{i}{2}\left(
\begin{array}{cc}
0 & 0 \\
0 & -\tau^T_{\mathrm{\,i}^\prime} \\
\end{array}
\right)\,.
\end{equation}
Substituting \p{M-1-a} and \p{eta-sigma1} into \p{Mpm-eta-inv}, one finds
\begin{equation}
\label{Si-pm-ch}
S^{(+)}_{\mathrm{\,i}} = -\frac{i}{2}\,\rho^i_I (\tau_{\mathrm{\,i}})_i{}^j\xi_j^I\,,\qquad
S^{(-)}_{\mathrm{\,i}^\prime} = -\frac{i}{2}\,\rho^{\underline{i}}_I (\tau_{\mathrm{\,i}^\prime})_{\underline{i}}{}^{\underline{j}}\xi_{\underline{j}}^I\,.
\end{equation}
Thus, the generators $S^{(+)}_{\mathrm{\,i}}$ are built using the canonical pairs $(\xi_i^I,\rho^j_J)$ from \p{pi-2-2},
while the generators $S^{(-)}_{\mathrm{\,i}^\prime}$ are built using the other canonical pairs $(\xi_{\underline{i}}^I,\rho^{\underline{j}}_J)$.

Now using the matrix expressions
\begin{equation}
\label{eta-sigma2}
(\tilde\sigma^{+\hat a})^{\alpha}{}_{\beta}=
\frac{i}{\sqrt{2}}\,
\left(
\begin{array}{cc}
0 & 0 \\
-\tau_{\hat a}^T & 0 \\
\end{array}
\right)
\,,\quad
(\tilde\sigma^{-\hat a})^{\alpha}{}_{\beta}=
\frac{i}{\sqrt{2}}\,
\left(
\begin{array}{cc}
0 & \tau_{\hat a}^T \\
0 & 0 \\
\end{array}
\right)\,,\ \ \mbox{at} \ \ \hat a=1,2,3\,,
\end{equation}
\begin{equation}
\label{eta-sigma3}
(\tilde\sigma^{+4})^{\alpha}{}_{\beta}=
\frac{-1}{\sqrt{2}}\,
\left(
\begin{array}{cc}
0 & 0 \\
1_2 & 0 \\
\end{array}
\right)
\,,\quad
(\tilde\sigma^{-4})^{\alpha}{}_{\beta}=
\frac{-1}{\sqrt{2}}\,
\left(
\begin{array}{cc}
0 & 1_2 \\
0 & 0 \\
\end{array}
\right)\,,\quad
(\tilde\sigma^{+-})^{\alpha}{}_{\beta}=
\frac{1}{2}\,
\left(
\begin{array}{cc}
-1_2 & 0 \\
0 & 1_2 \\
\end{array}
\right)
\end{equation}
obtained from \p{sigma-a}, \p{sigma-a-r}, \p{sigma-m}, and expansion \p{pi-2-2}, we find the spin part of the remaining
Lorentz group generators \p{M-1}:
\begin{equation}
\label{Si-pm-ch-1}
S^{+\hat a}=
-\frac{i}{\sqrt{2}}\,\rho^i_I(\tau_{\hat a})_{i}{}^{\underline{j}}\xi_{\underline{j}}^I
\,,\quad
S^{-\hat a}=
\frac{i}{\sqrt{2}}\,\rho^{\underline{i}}_I(\tau_{\hat a})_{\underline{i}}{}^{j}\xi_{j}^I
\,,\ \ \mbox{at} \ \ \hat a=1,2,3\,,
\end{equation}
\begin{equation}
\label{Si-pm-ch-2}
S^{+4}=
-\frac{1}{\sqrt{2}}\,
\rho^i_I \delta_{i}{}^{\underline{j}}\xi_{\underline{j}}^I
\,,\qquad
S^{-4}=
-\frac{1}{\sqrt{2}}\,
\rho^{\underline{i}}_I \delta_{\underline{i}}{}^{j}\xi_{j}^I\,,\qquad
S^{+-}=
-\frac{1}{2}\left(\xi_{i}^I \rho^{i}_I -  \xi_{\underline{i}}^I \rho^{\underline{i}}_I\right)\,.
\end{equation}

The found expressions \p{Mpm-eta-summ}, \p{Si-pm-ch}, \p{Si-pm-ch-1}, \p{Si-pm-ch-2} are used in Sect.\,4 to construct
the spin part of the Lorentz algebra generators in the bi-harmonic space.

\end{document}